\documentclass[sigconf,nonacm]{acmart}

\AtBeginDocument{%
  \providecommand\BibTeX{{%
    \normalfont B\kern-0.5em{\scshape i\kern-0.25em b}\kern-0.8em\TeX}}}

\setcopyright{acmlicensed}
\copyrightyear{2026}
\acmYear{2026}
\acmDOI{XXXXXXX.XXXXXXX}

\usepackage{url}

\usepackage{caption}
\usepackage[all]{nowidow}
\usepackage{wrapfig}
\usepackage{arydshln}
\usepackage{multirow}
\usepackage{tabularx}
\usepackage{array}
\usepackage{ragged2e}
\usepackage{makecell}
\newcolumntype{L}[1]{>{\raggedright\let\newline\\\arraybackslash\hspace{0pt}}m{#1}}
\newcolumntype{C}[1]{>{\centering\let\newline\\\arraybackslash\hspace{0pt}}m{#1}}
\newcolumntype{R}[1]{>{\raggedleft\let\newline\\\arraybackslash\hspace{0pt}}m{#1}}

\usepackage{wrapfig,lipsum,booktabs} 

\def\authnotes{1}
\newcounter{notectr}[section]
\newcommand{\thenote}{\thesubsection.\arabic{notectr}\refstepcounter{notectr}}

\newcommand{\note}[2]{$\ll$#1~\thenote: #2$\gg$}
\newcommand{\cnote}[1]{\ifnum\authnotes=1 \textcolor{blue}{\note{Comment:}{#1}}\fi}

\newcolumntype{L}[1]{>{\RaggedRight\arraybackslash}p{#1}}
\newcolumntype{Y}{>{\RaggedRight\arraybackslash}X}

\begin{document}




\title[A ``Good'' Family Life Through Energy Practices]{``You Can't Just Automate It'': Negotiating and Sustaining a ``Good'' Family Life Through Energy Practices}

\author{Yang Hong}
\authornote{These authors contributed equally to this work.}
\affiliation{%
  \institution{University of Illinois Urbana-Champaign}
  \city{Champaign}
  \state{Illinois}
  \country{United States}}
\email{yangh9@illinois.edu}

\author{Ying-Yu Chen}
\authornotemark[1]
\affiliation{%
  \institution{National Cheng Kung University}
  \city{Tainan}
  \country{Taiwan}}
\email{yingyuchen@gs.ncku.edu.tw}

\author{Wei-Chien Chang}
\affiliation{%
  \institution{Corp. Innovation \& Research, Compal Electronics}
  \city{Taipei City}
  \country{Taiwan}}
\email{Vivi\_Chang@compal.com}

\author{Yu-Hsin Chou}
\affiliation{%
  \institution{ University of Washington}
  \city{Seattle}
  \state{Washington}
  \country{United States}}
\email{ychou3@uw.edu}

\author{Sharifa Sultana}
\affiliation{%
  \institution{University of Illinois Urbana-Champaign}
  \city{Champaign}
  \state{Illinois}
  \country{United States}}
\email{sharifas@illinois.edu}

\renewcommand{\shortauthors}{Hong et al.}

\begin{abstract}
This study examines how Taiwanese parent-child families negotiate a ``good'' family life through everyday energy use and imagine future smart homes that support it. We conducted in-home interviews and co-design sessions with 21 families, including 46 parents and children. We found that families pursued a ``good'' life through energy practices shaped by thrift, comfort, care, safety, and enjoyment. These arrangements were continually adapted and responded to changing bodies, schedules, people, and infrastructures. This adaptive work was unevenly distributed, which in turn shaped different smart-home imaginaries. Drawing on the lens of Nearby and adversarial design, we conceptualize adaptation as situated sociotechnical work through which families continually rework energy arrangements. We further distinguish \textit{collective goods} from \textit{plural and contestable goods} to show why family IoT must support shared values while preserving opportunities to question and revise household arrangements. We offer theoretical and design directions for more adaptive, participatory, and contestable family IoT.
\end{abstract}







\settopmatter{printfolios=true}

\maketitle

\section{Introduction}

What counts as necessary, excessive, or sustainable household energy use is not universal. It is shaped by climate, housing, and cultural expectation. European households may treat air-conditioning as an indulgence; Taiwanese families may treat heating the same way, reaching for a blanket instead. Smart-home sustainability systems, which promise to reduce consumption through sensing, feedback, and automation \cite{uncovering2013Schwartz, alan2016toohot, jensen2018desirable, 3P2019, gyllensward2006visualizing, yang2014}, tend to flatten this variation, with an embedded assumption about what a well-run home looks like --- efficient, predictable, and automatic. Critical HCI scholarship has long named the costs of this assumption, through figures like \textit{Resource Man} and the \textit{Smart Wife} \cite{resourceman, smartwife, VERKADE201736}, and through arguments that persuasive and behavior-change approaches quietly convert sustainability into an individual optimization problem, stripped of the social and political conditions that make consumption meaningful in the first place \cite{dourish2011divining, fogg2002persuasive, persuasive2024, persuasivenarrow}. DiSalvo pushes this further: instead of design to resolve conflict into consensus, he advocates adversarial design that stages contest and keeps visible whose interests and values are actually at stake \cite{adversarialdesign}.

This critique has mostly been made at the level of the household-as-user. Less is known about what happens when the household is composed of multiple users with unequal power. Family HCI research shows access and authority over technology are unevenly distributed. Men more often configure technology, women more often coordinate its day-to-day use \cite{3P2019, geeng2019control, strengers2018aesthetic, rode2018gender, strengers2011designing}, children sit furthest from control but are not passive \cite{fell2014children, WANG2025103439}. A thermostat's schedule protects somebody's comfort \cite{alan2016toohot, yang2013learning, yang2014}; a locked configuration enacts somebody's authority \cite{geeng2019control, xue2024control}. Smart-home systems that encode one set of preferences give this everyday contest almost nowhere to go. This paper extends this critique of automation into the parent-child family by asking who contests household energy use and who has to give way.

We take up this question in Taiwan, where energy conservation is strongly bound up with intergenerational thrift and the everyday injunction to ``not waste.'' Carbon targets and household bills form only part of this picture. Running an air conditioner, dehumidifier, or light in these homes carries meanings beyond consumption. It can support comfort, mold prevention, care for a child or grandparent, and lessons about responsible family membership. But this ``good'' family life cannot be fixed in advance \cite{mealtime, goodsleep}. An AC scheduled off at 2 AM is efficient until a child is still awake; a light left on is wasteful until a child is afraid of the dark. These are not deviations from an optimal household; they are what family life is made of.

We draw on Xiang's concept of the \textit{Nearby} \cite{xiang2021nearby}, which directs attention to the people, differences, and repeated encounters that organize everyday life. We use this lens to conceptualize \textit{adaptation} as the situated work where families notice changing circumstances and rework energy arrangements accordingly. Where existing smart-home imaginaries assume a good life can be specified once and stabilized through automation, we show families instead sustain it through continual adaptation, unevenly distributed across parents and children. We read this distribution through DiSalvo's adversarial design \cite{adversarialdesign}. Rather than seeing the house as a unit for automation to harmonize, we examine it as a small, ongoing arena of contest over whose comfort, labor, and authority a given energy arrangement serves.

To study this, we conducted a qualitative study with 21 Taiwanese parent-child families (46 parents and children in total), combining semi-structured interviews with co-design sessions on home energy technologies. We ask:

\begin{itemize}
    \item \textbf{RQ1:} How do parent-child families make sense of “good” family life through everyday home energy practices in Taiwan? 
    \item \textbf{RQ2:} How do families adapt energy practices and smart-home arrangements as household needs and circumstances change? 
    \item \textbf{RQ3:} How do family members' experiences of this adaptive work shape their imaginaries of future smart-home technologies? 
\end{itemize}

We find that energy conservation was embedded in interwoven, situated judgments about thrift, comfort, care, safety, and household responsibility. Sustaining these judgments required continual adaptive work that was unevenly distributed across family members. Fathers more often configured technology, mothers more often adjusted it to changing conditions, and children negotiated and resisted from the margins. These different positions produced different imaginaries of future smart homes, ranging from centralized automation to flexible designs that accommodate exceptions.

We make three contributions to sustainable and family-centered HCI. 
\textbf{First}, we provide an empirical account of how families negotiate energy consumption through shared and contested values that give culturally situated meaning to what constitutes a ``good'' family life through energy use. 
\textbf{Second}, drawing on the \textit{Nearby} and care theory, we conceptualize \textit{adaptation} as the situated care work that families continually rework energy arrangements as relations among people, devices, routines, and conditions change. We further show that this work is unevenly distributed across gender and generation, and that smart-home automation can reconfigure rather than remove it. 
\textbf{Third}, we contribute design implications for family IoT by distinguishing between plural and contestable good and collective good. We argue that smart-home systems should keep plural good visible and revisable, support the intergenerational articulation of collective good such as thrift, and remain responsive to the adaptive care work through which both are sustained in everyday life.
\section{Related Work}

\subsection{Care and Everyday Life in Smart Homes}
\subsubsection{From Automation to Relational Care} HCI research has increasingly moved away from understanding smart homes simply as collections of automated devices toward examining how technologies become embedded in everyday relations of care, domestic labor, and dwelling. Smart-home technologies have long been envisioned as infrastructures for supporting security, comfort, household management, entertainment, and independent living \cite{duque2022troubleshooting,strengers2022marvelous}. More recent research has emphasized that these forms of technological assistance are relational rather than purely functional. Maalsen, for example, argues that care enacted through algorithmic and automated systems is not necessarily one-directional: smart technologies can shape how household members themselves notice needs, respond to others, and perform care \cite{maalsen2023algorithmic}. Related work on smart-home care ecologies further broadens attention from individual devices to the relationships among technologies, residents, dwelling conditions, and wider housing infrastructures \cite{liu2026home}.

\subsubsection{Invisible Domestic Labor, Repair, and Everyday Maintenance}
This perspective also complicates longstanding promises that automation will simply reduce domestic work. Feminist scholarship has shown that household technologies can redistribute or even increase the invisible labor required to make domestic systems function, rather than eliminate it. Research on future care work similarly asks not only whether care can be automated, but what kinds of care work should be automated, for whom, and with what consequences \cite{wu2024collective,pillai2025speculating}. Studies of smart homes reveal comparable dynamics in everyday practice. Devices may fail to align with routines, respond unexpectedly, require troubleshooting, or demand workarounds and ongoing maintenance \cite{duque2022troubleshooting,liu2026humor}. Rather than treating these moments only as technological breakdowns, researchers have shown how household members make such misalignments livable through experimentation, repair, appropriation, curiosity, and even humor \cite{strengers2022marvelous,liu2026humor}. These studies challenge imaginaries of the smart home as seamless and autonomous by foregrounding the continuing human work required to sustain it.

\subsubsection{Care Ecologies, Dwelling, and Energy}
Care-oriented research also emphasizes that the home itself is not a neutral container for technological use. Housing arrangements, material conditions, and social relationships shape how technologies can be used and who benefits from them. Research on rented and shared smart homes shows that residents may live with technologies they did not choose, install, or fully control \cite{maalsen2023cheap}. Work on housing and care similarly demonstrates how care emerges through imperfect relationships among people, infrastructures, dwelling arrangements, and material conditions \cite{maalsen2026imperfect}. In parallel, research on family media practices has shown that household technologies become entangled with rules, authority, negotiation, and the ongoing construction of family relationships \cite{horst2010families}. Read together, this literature suggests that smart-home technologies should be understood as participants in broader ecologies of everyday care rather than as isolated tools that act upon individual users.

However, care-oriented smart-home research has focused primarily on housing, aging, domestic labor, wellbeing, and future care technologies. Less attention has been paid to how care becomes intertwined with everyday energy use and sustainability practices within families. Sustainable HCI has extensively studied household energy consumption, while research on smart-home care has shown how technologies become embedded in relational and material forms of everyday life; these conversations have remained comparatively separate. Our work connects them by examining energy consumption itself as part of family care: decisions about cooling, lighting, dehumidification, comfort, safety, and avoiding waste are simultaneously decisions about how family members care for one another and maintain a desirable domestic life. We build on this literature to examine how such arrangements must be continually adjusted as people, routines, technologies, spaces, and environmental conditions change.

\subsection{Multi-User Smart Homes and Family Energy Practices}
\subsubsection{From Energy Feedback to Social Practice}
Research on sustainable HCI initially approached household energy consumption largely through behavioral intervention. Drawing on persuasive technology and eco-feedback, early systems provided real-time consumption data, ambient displays, or behavioral prompts with the expectation that greater awareness would encourage individuals to reduce energy use \cite{fogg2002persuasive,gyllensward2006visualizing,mccalley2006from,midden2009using}. Subsequent research challenged this assumption. Studies found that users often returned to existing routines over time and that treating consumption as an individual, rational decision overlooked the habitual, emotional, social, and collective organization of domestic life \cite{strengers2011designing,pierce2010materializing,brynjarsdottir2012sustainably}.

This critique led sustainable HCI researchers to examine energy use as a socially embedded practice. Everyday activities associated with comfort, cleanliness, convenience, care, and household roles may take priority over abstract sustainability goals \cite{strengers2011designing,pierce2012beyond}. Pierce and Paulos, for example, describe some domestic practices as effectively ``non-negotiable'' when residents are unwilling to abandon them despite their environmental consequences \cite{pierce2012beyond}. Similarly, studies of family engagement with smart-home technologies show that comfort, convenience, aesthetics, and desirability can outweigh sustainability as explicit motivations for adoption \cite{jensen2018desirable}. More recent sustainable-HCI approaches consequently argue for supporting flexible and situated household practices rather than designing only around predetermined energy-saving behaviors \cite{asgeirsdottir2023energy,guizzardi2025sustainability}.

\subsubsection{Uneven Control in Multi-User Smart Homes}
Recognizing households as multi-user environments further complicates assumptions about individual control. Smart-home research has shown that access, configuration, and decision-making are often unevenly distributed among household members \cite{pins2021alexa,geeng2019control,koshy2021passenger}. Koshy et al. distinguish between pilot users, who install technologies and determine settings and routines, and passenger users, who must live with those configurations despite having less knowledge of or access to system controls \cite{koshy2021passenger}. Such arrangements often privilege technically confident adults and can reproduce gendered divisions of domestic authority, with men more commonly occupying configuration roles while women perform substantial everyday household work \cite{3P2019}. Children are frequently positioned even further from technical control, although prior research demonstrates that they can actively influence household energy practices when given meaningful responsibility and access \cite{fell2014children}.

\subsubsection{Negotiating Energy, Comfort, and Family Authority}
The above-mentioned asymmetries matter because household energy use is rarely determined by a single user. Families must continually coordinate different preferences for temperature, lighting, device use, comfort, schedules, and expenditure. Yet many smart-home systems encode predefined values of efficiency and automation while providing limited support for dialogue, disagreement, or changing circumstances among cohabitants \cite{xue2024control,yang2014}. Energy-feedback systems can similarly individualize responsibility by presenting consumption data without supporting the relational processes through which families collectively interpret and act on it \cite{abrahamse2009energy}. Research on automated heating controls further shows how algorithmically determined settings can constrain residents’ abilities to express changing bodily comfort and social routines \cite{alan2016toohot,yang2014}. Thus, automation may reduce some visible coordination while simultaneously creating new questions about whose preferences are encoded, who can modify them, and who must adapt when they no longer fit everyday life.

Research on families and new media likewise shows that household rules are continually created, negotiated, bent, and broken as parents and children establish authority and family identity through technology use \cite{horst2010families}. However, less is known about how these parent–child dynamics intersect specifically with everyday energy practices and smart-home arrangements, particularly outside Western contexts. Children are often discussed as users to be educated about conservation, while parents are treated as relatively coherent household decision-makers. This framing can obscure differences among fathers, mothers, children, and grandparents in how they understand waste, comfort, care, responsibility, and technological control. Our study builds on this body of work by examining Taiwanese parent-child families as multi-user energy ecologies in which sustainability is continually negotiated and adjusted. We examine how family members make sense of ``good'' energy use, how they coordinate and adapt practices as household needs and circumstances change, and how uneven experiences of configuration and adjustment shape their imaginaries of future smart homes.
\section{Methods}

We report qualitative findings from an in-home study with 21 Taiwanese families (46 participants: 23 parents and 23 children aged 7 to 12) over two years (Table~\ref{tab:participant-demographics}). We combined semi-structured interviews, home tours, visual mapping, and co-design activities to examine parents' and children's energy practices and negotiations in everyday life.

\begin{table*}[t]
  \caption{Participant demographics and energy-related devices in the home (All participant and family names are pseudonyms).}
  \label{tab:participant-demographics}

  \footnotesize
  \setlength{\tabcolsep}{3.5pt}
  \renewcommand{\arraystretch}{1.08}

  \begin{tabularx}{\textwidth}{
    @{}
    L{1.65cm}
    L{3.25cm}
    L{2.35cm}
    L{2.55cm}
    Y
    @{}
  }
        \toprule
    \textbf{Family}
      & \multicolumn{2}{c}{\textbf{Parent}}
      & \textbf{Child}
      & \multirow{2}{=}{\centering\textbf{Energy-related Devices}} \\
    \cmidrule(lr){1-1}
    \cmidrule(lr){2-3}
    \cmidrule(lr){4-4}
    
    \textbf{ID / Name}
      & \textbf{Name (Age/Gender)}
      & \textbf{Occupation}
      & \textbf{Name (Age/Gender)}
      & {} \\
    \midrule
    
    F1--Smith
      & Ashley (30--39/Female) \newline
        Brandon (P1-b/30--39/Male)
      & Teachers \newline
        Professionals
      & Emma (7--12/Female)
      & Voice Assistant (Xiaomi), Smart Lighting (Xiaomi),
        Automatic Watering System, Automatic Door Closing System
        (Xiaomi) \\
    \addlinespace[2pt]

    F2--Johnson
      & Jennifer (40--49/Female)
      & Household management
      & Liam (7--12/Male)
      & Voice Assistant (Google),
        \textit{Smart Lamps}\textsuperscript{*} \\
    \addlinespace[2pt]

    F3--Williams
      & Michael (40--49/Male)
      & Service Workers and Salespersons
      & Olivia (7--12/Female)
      & Voice Assistant (Xiaomi),
        \textit{Auto Flush Toilet}\textsuperscript{*} \\
    \addlinespace[2pt]

    F4--Brown
      & Brittany (30--39/Female)
      & Service Workers and Salespersons
      & Ava (7--12/Female)
      & \textit{Intelligent Lighting}\textsuperscript{*},
        \textit{Solar Thermal Storage System}\textsuperscript{*} \\
    \addlinespace[2pt]

    F5--Jones
      & Samantha (30--39/Female)
      & Household management
      & Noah (7--12/Male)
      & Intelligent Water Tap, Intelligent Outdoor Lighting System \\
    \addlinespace[2pt]

    F6--Garcia
      & Lauren (30--39/Female)
      & Professionals
      & Sophia (7--12/Female)
      & Smart Refrigerator (Panasonic),
        \textit{Solar Power System}\textsuperscript{*} \\
    \addlinespace[2pt]

    F7--Miller
      & Christopher (40--49/Male)
      & Professionals
      & Jackson (7--12/Male)
      & Voice Assistant (Google), Smart Lighting (Google) \\
    \addlinespace[2pt]

    F8--Davis
      & David (40--49/Male)
      & Professionals
      & Lucas (7--12/Male)
      & Voice Assistant (Apple), Smart Lighting,
        Smart Energy Box (Amazon), Smart Meter \\
    \addlinespace[2pt]

    F9--Ross
      & Matthew (40--49/Male)
      & Professionals
      & Aiden (7--12/Male)
      & Voice Assistant (Xiaomi), Smart Lamps (Xiaomi) \\
    \addlinespace[2pt]

    F10--Martin
      & Joshua (40--49/Male)
      & Professionals
      & Isabella (7--12/Female)
      & Voice Assistant (Amazon), Smart Lamps \\
    \addlinespace[2pt]

    F11--Hall
      & Amanda (30--39/Female)
      & Professionals
      & Ethan (7--12/Male)
      & \textit{Smart Lamps}\textsuperscript{*},
        Smart Monitor, Robot Vacuum \\
    \addlinespace[2pt]

    F12--Lopez
      & Melissa (40--49/Female)
      & Household management
      & Mason (7--12/Male) \newline
        Logan (7--12/Male)
      & \textit{Smart Lighting}\textsuperscript{*},
        Robot Vacuum, Smart Air Purifier \\
    \addlinespace[2pt]

    F13--Clark
      & Nicole (40--49/Female)
      & Professionals
      & Mia (7--12/Female)
      & \textit{Smart Water Heater}\textsuperscript{*} \\
    \addlinespace[2pt]

    F14--Wilson
      & Jessica (30--39/Female)
      & Professionals
      & Charlotte (7--12/Female)
      & Smart Monitor,
        \textit{Smart Water Heater}\textsuperscript{*} \\
    \addlinespace[2pt]

    F15--Anderson
      & Stephanie (40--49/Female)
      & Professionals
      & Elijah (7--12/Male)
      & Voice Assistant (Alexa), Robot Vacuum \\
    \addlinespace[2pt]

    F16--Thomas
      & Tyler (30--39/Male)
      & Professionals
      & Amelia (7--12/Female)
      & Voice Assistant (Xiaomi), Automatic Door Closing System
        (Xiaomi), Smart Lamps \\
    \addlinespace[2pt]

    F17--Taylor
      & Rachel (30--39/Female)
      & Professionals
      & James (7--12/Male)
      & Smart Monitor, Smart Air Purifier \\
    \addlinespace[2pt]

    F18--Moore
      & Heather (40--49/Female)
      & Household management
      & Harper (7--12/Female)
      & \textit{Smart Water Heater}\textsuperscript{*}, Robot Vacuum \\
    \addlinespace[2pt]

    F19--Jackson
      & Angela (40--49/Female)
      & Household management
      & Benjamin (7--12/Male)
      & Voice Assistant (Xiaomi, Xiaodu, Tmall) \\
    \addlinespace[2pt]

    F20--Martin
      & Kayla (20--29/Female)
      & Professionals
      & Evelyn (7--12/Female)
      & Smart Monitor,
        \textit{Smart Lamps}\textsuperscript{*} \\
    \addlinespace[2pt]

    F21--Lee
      & Kevin (30--39/Male) \newline
        Megan (30--39/Female)
      & Professionals \newline
        Professionals
      & Abigail (7--12/Female) \newline
        Henry (7--12/Male)
      & Voice Assistant (Google Home), Smart Lighting,
        Smart Socket, Smart Monitor \\
    \bottomrule
  \end{tabularx}

  \vspace{3pt}
  \parbox{\textwidth}{%
    \footnotesize
    \textsuperscript{*}Appliances that participants identified as
    smart-home IoTs but that cannot interconnect with other devices
    via the Internet.
  }
\end{table*}

\subsection{Participants}

We recruited families with at least one child aged 7 to 12 and household experience with smart-home or Internet of Things (IoT) devices related to water, electricity, or gas use. During recruitment, families reported the smart appliances they used for household energy management. We initially used smart-home and IoT devices to refer to technologies that supported automation, remote operation, scheduling, or connectivity across devices, such as app-controlled air conditioners and voice assistants connected to lighting.

During the study, we found that participants used ``smart'' more broadly. Some described appliances such as thermostatic water heaters and air conditioners as smart even when they could not connect with other devices. Others identified smart devices, such as monitors or voice assistants used for entertainment, that were not directly involved in energy management. Table~\ref{tab:participant-demographics} therefore reports smart devices as identified by participants. For our analysis, we used the broader category of \textit{energy-related technologies} to include both connected and non-connected appliances that shaped families' everyday energy practices.

\subsection{Study Procedures}

Each family participated in an in-home session lasting approximately 2.5 to 3 hours, with at least one parent and one child participating from each household. Before data collection, we explained the study procedures and obtained informed consent from participating parents and children. We provided children with a child-friendly consent form that included Mandarin phonetic symbols to support comprehension and reminded participants that they could withdraw at any time. The study protocol was approved by the university's Institutional Review Board (IRB). With participants' permission, we audio-recorded interviews and photographed or video-recorded energy-related devices as participants explained how family members used them.

\begin{figure*}[t!]
    \centering
    \includegraphics[width=0.9\textwidth]{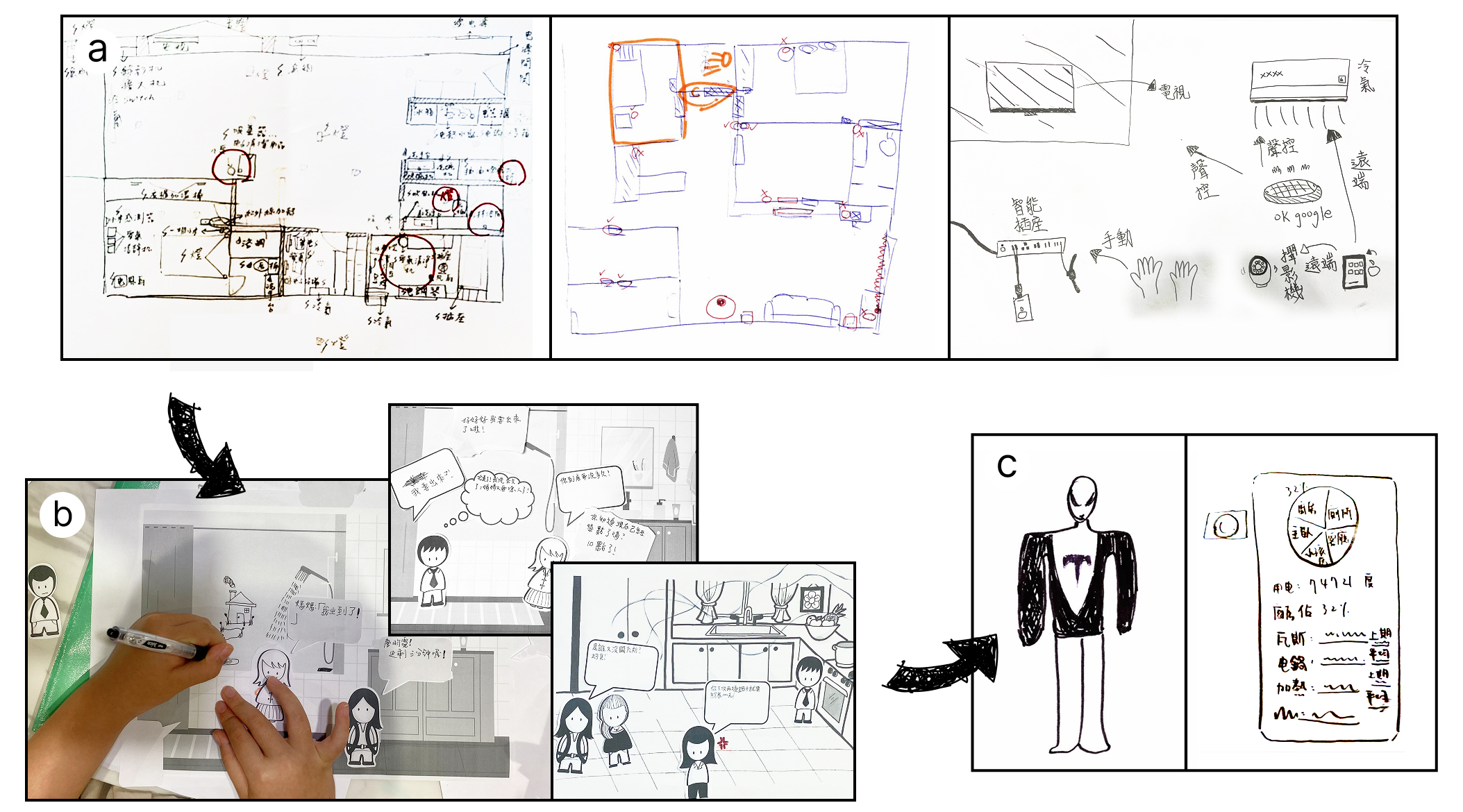}
    \caption{Study activities following the room tour. (a) Parents mapped energy-related devices on floor plans, while children drew devices they used in everyday life; families later marked locations where energy-related tensions occurred. (b) Participants reconstructed a specific episode of tension using prepared storyboard materials. (c) Parents and children co-designed technologies in response to these episodes. Shown are Jackson's reminder robot and Ashley's centralized interface for monitoring and controlling household energy use.}
    \label{fig:process}
\end{figure*}

The in-home session consisted of four parts. \textbf{First}, participants guided the research team through the home and identified where water, electricity, and gas were used, including energy use involving smart and conventional appliances. The room tour grounded subsequent questions in the devices, spaces, and routines participants used in everyday life.

\textbf{Second}, parents drew simple floor plans showing the locations of energy-related devices, while children drew the devices they personally used (Figure~\ref{fig:process}a). These materials captured parents' and children's perspectives separately and were used as prompts in the following interviews. \textbf{Third}, we interviewed parents and children individually about their everyday energy practices and experiences negotiating energy conservation with other family members.

\textbf{Finally}, parents and children participated together in a co-design activity. They revisited their maps to identify locations where energy negotiations were difficult, reconstructed a specific episode using prepared storyboards, and then imagined technologies that might support the family in that situation (Figure~\ref{fig:process}b--c). Moving from a concrete household episode to a future design grounded participants' speculation in tensions they had experienced in everyday life.

\subsection{Data Analysis}

All interviews were conducted and transcribed in Mandarin and analyzed using reflexive thematic analysis \cite{braun2006thematic, Byrne2022, Braun2019}. The research team translated selected excerpts into English, with back-translation used to check consistency with the original Mandarin. We first read the transcripts repeatedly to become familiar with each family's practices and interactions. We then manually coded the data, using the research questions to orient the analysis while developing codes inductively from participants' accounts. Three authors participated in the coding and met regularly to compare interpretations, refine codes, and develop candidate themes. These discussions were used to deepen our interpretation of the data. We repeatedly reviewed the themes against the transcripts and refined their boundaries as the analysis progressed. The resulting themes captured how family members understood desirable energy use, negotiated and adapted household practices, and imagined future energy-related technologies. The final codebook can be found in the Appendix.
\section{Taiwanese Smart Home and Energy Use}


Taiwanese household energy use takes place within a policy environment that combines energy-saving programs with relatively low and stable utility prices. The government has promoted smart meters, energy-efficient appliances, and subsidies for replacing older air conditioners and refrigerators as part of its digitalization and energy-saving agenda \cite{executiveyuan2017smartcity,executiveyuan2024energysaving}. At the same time, residential electricity prices are managed through a government tariff review process that considers supply costs, inflation, household burden, and broader political-economic conditions. Residential rates have often remained below the cost of supplying electricity \cite{moea2025tariff}. Water prices have likewise remained low and largely unchanged for more than two decades \cite{taiwanwater2025tariff}. This pricing context makes it important to look beyond utility bills when considering why Taiwanese families conserve energy.

Energy conservation also carries cultural and familial meanings in Taiwan. Chiu shows that ``energy conservation'' resonates with cultural values of frugality and the preciousness of resources, while ``carbon reduction'' can be more abstract and contested \cite{chiu2013tensions}. Family relationships provide another important context. Filial piety, rooted in Confucian family ethics, continues to shape ideas about parental authority and children's responsibilities. Research with Taiwanese adolescents links authoritarian filial piety with stronger beliefs in parental authority and obedience \cite{liu2013autonomy}. Household labor also remains gendered, with women continuing to carry greater responsibility for routine domestic work in many Taiwanese families \cite{chung2022gendering}. Women also tend to take on more care work and responsibility for childrearing and education \cite{goodsleep, TWwomencare}. Energy practices therefore sit within everyday processes of family education, authority, and care.

The smart-home environment in our participating households was also fragmented. Families used devices from ecosystems such as Google Home, Amazon Alexa, and Xiaomi, together with individual smart appliances made by Taiwanese and other manufacturers. These products often required separate apps and offered limited integration across brands. Some families therefore assembled their own connections through smart switches, sensors, voice control, and one-button scenes. For example, in the Smith family (F1), Brandon linked several Xiaomi devices through one-button controls. Similar setups appeared in families where husbands were smart-home enthusiasts (F8, F16, F21), who purchased devices from different regions and configured automated connections among them. These policy, cultural, and technical conditions form the setting for our study. 
\section{Findings}

Our findings show how Taiwanese families pursue and sustain a \textit{good} family life through everyday energy use. We first examine how families define appropriate energy use through values such as thrift, comfort, care, safety, and enjoyment. We then show how these arrangements require continual \textit{adaptation} as family life and household conditions change. Finally, our co-design sessions show how family members' different experiences of this adaptive work shaped their imaginaries of future smart-home technologies.

\begin{figure*}[t!]
    \centering
    \includegraphics[width=\textwidth]{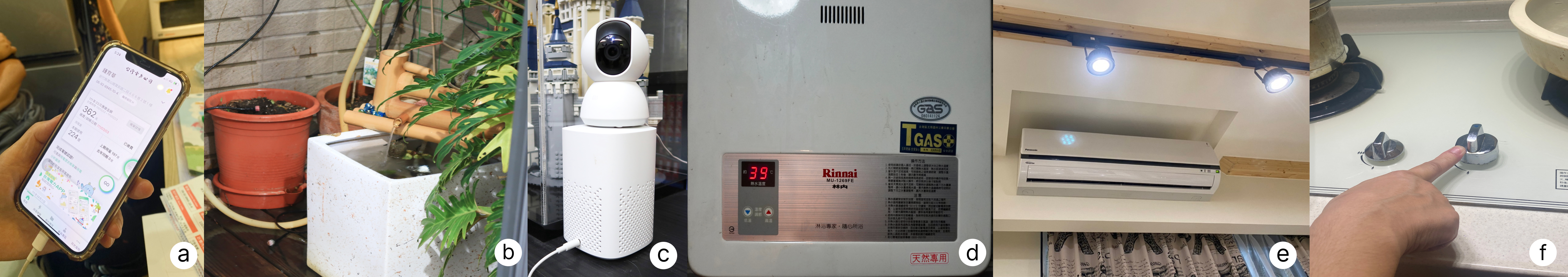}
    \caption{During room tours, participants showed us energy-related devices and apps in their homes: (a) a mobile app for paying electricity bills; (b) an automatic watering system on the balcony; (c) Xiao Ai, a voice assistant used to control other IoT devices; (d) a water heater showing the current water temperature; (e) an air conditioner and lighting; and (f) a kitchen gas stove.}
    \label{fig:roomtour}
\end{figure*}

\subsection{Constructing a \textit{Good} Family Life Through Energy Use}

Family members held different visions of a \textit{good} life, which shaped different goals for energy use. These included carrying forward thrift as a traditional virtue, pursuing bodily comfort, caring for the home and other family members, and making room for fun.

\subsubsection{Thrift and Avoiding Waste}

Thrift in energy use, often expressed as \textit{``not wasting''}, was repeatedly mentioned across all participating families. Parents described similar practices, such as turning off unused lights and appliances, reusing water, limiting air-conditioner(AC) use, and shortening showers. Ten parents traced these practices to their own upbringing, and eighteen described thrift as a family virtue passed across generations.

In the Smith family (F1), Brandon similarly told us that he had developed energy-saving habits from childhood and continued to see them as a traditional virtue even as the living conditions had improved. Nicole (Clark, F13) likewise described turning off lights when leaving a room as a \textit{``good habit''} learned in childhood: \textit{``Our generation was taught from a young age by our parents not to waste and to turn off the lights when we leave.''} Parents carried similar expectations forward to their children. Melissa (Lopez, F12) explained: 

\begin{quote}
    \textit{``Since we were young, our parents would remind us when the lights were left on. Now that we are adults, we tell our children the same thing. They need to be responsible for their own room.''} 
\end{quote}

Small energy-saving practices thus became ways of teaching children responsible participation in family life. Many children also understood thrift through concrete actions. Olivia (Williams, F3) explained: \textit{``Saving energy means turning them off when you leave a room.''} Utility bills also provided a concrete way to teach thrift. Lauren (Garcia, F6) connected household costs with responsible resource use. She used bills to teach her child about thrift:

\begin{quote}
    \textit{``I tell her, `You are wasting our electricity. These resources do not come easily, and Mom and Dad still have to pay the electricity and water bills. We should save energy together so there is enough for people who need it. When the power or water goes out, it is very inconvenient, so we should not waste it.' ''}
\end{quote}

While at the same time, a majority of parents stated that electricity or water in Taiwan is quite affordable and stable, as described by Joshua (Martin, F10), \textit{``resources like these are relatively cheap in Taiwan.''} David (Davis, F8) recalled that he once forgot to turn off LED lights for long periods, only making little difference to the bill. Nicole explained this as a generational shift in the meaning of thrift:

\begin{quote}
    \textit{``Older generations thought wasting electricity meant wasting money. Their concern was entirely about cost, not saving energy. Now that people are better off financially, we think of it more as conserving energy. That is where the idea has changed.''}
\end{quote}

As utility expenses themselves became less salient, \textit{``not wasting''} increasingly emphasized valuing resources and avoiding unnecessary energy use. Michael (Williams, F3) summarized this principle as: \textit{``Just use energy normally and don't waste it.''} He distinguished ordinary use from consumption that continued without serving anyone. Nine other parents also emphasized \textit{``normal use.''} Joshua explained that \textit{``Energy use was reasonable when someone was actually benefiting from it, so leaving the AC on after leaving a room was unnecessary.''} Reducing such extra consumption therefore became a common basis for household energy rules.

\subsubsection{Comfort}

Earlier, families framed the overall goal of household energy use as \textit{not wasting}. Yet what counted as normal use'' and what counted as waste was highly contextual. Comfort made this distinction especially difficult because family members could experience the same temperature differently. Nicole captured this problem: \textit{At the same temperature, one person feels hot and another feels cold.''} 
Fifteen parents mentioned 27$^\circ$C as a commonly recommended energy-efficient AC setting. Yet this temperature did not always fit family members' comfort needs. Angela in the Jackson family (F19) shared her energy-saving strategy for AC and her family's response:

\begin{quote}
\textit{``I saw energy-saving experts on TV recommend this. I also think this saves the most energy: first set the AC very low so the room cools down quickly, then raise it back to 27 degrees and keep it there instead of turning it on and off [...] I think my children may still feel hot. They go stay in their dad's room because he keeps it much colder, maybe around 22 degrees.''}
\end{quote}

The tensions between energy-efficient temperature and other family members' bodily comfort also emerged in the Jones family (F5). Samantha described her husband as needing cooling wherever he went, sometimes turning on the downstairs AC while mopping and preferring much of the home around 23$^\circ$C. She and her children jokingly called him a \textit{portable air conditioner}:

\begin{quote}
\textit{``Dad literally needs a `portable AC.' Wherever he goes, the AC goes on. In our house, wherever Dad is, the AC is unlimited [...] He wants it at 22 degrees. But sometimes that is freezing for the kids.''}
\end{quote}

Rather than setting another fixed temperature, Samantha checked the children before bed and adjusted the AC according to how they were sleeping: \textit{``I go look at them before bed. If they look really cold and are wrapped tightly in their blankets, I adjust the AC back toward room temperature, maybe 26 degrees.''}

Repeatedly adjusting the setting around different family members' needs became one way to temporarily ease these tensions. The Smith family (F1) encountered a similar disagreement around hot water. Brandon, an engineer and smart-home enthusiast, capped the gas water heater at 42$^\circ$C. Given the distance between the balcony heater and the bathrooms and heat loss through the pipes, he saw heating water further only to mix it with cold water as wasteful: \textit{``If I set the water very hot, by the time it reaches the farther bathroom I still have to add cold water before I can shower. To me, that is too wasteful.''} His wife, Ashley, experienced the same setting differently. She preferred hotter water and traced this difference to the habits she and Brandon had learned in their families growing up:

\begin{quote}
\textit{``My child and my husband shower with cooler water. He normally sets the maximum water temperature to 42 degrees. In the family I grew up in, we would not lower the maximum temperature of the gas water heater. But my husband always lowers it. In winter, I raise it to 50 degrees myself. Then when he showers and finds the water too hot, he thinks it wastes energy and turns it back to 42 degrees. Sometimes I have already taken off my clothes and then realize the water is so cold. To me, that temperature feels like cold water [...] I feel uncomfortable. He does not respect the temperature I want. Finally, I gave in and got used to adjusting the water temperature myself before showering.''}
\end{quote}

Their repeated adjustments show how the same energy setting could embody competing ideas of what counted as an \textit{``appropriate temperature.''} These differences could also reflect broader cultural experiences. Stephanie in the Anderson family preferred wearing a sweater instead of heating the home, while her German husband was accustomed to warmer indoor temperatures:

\begin{quote}
\textit{``He is from Germany, and in winter people there keep their homes very warm. Even when it is below freezing outside, they might wear short sleeves indoors. They are just not used to wearing thick clothes at home.''}
\end{quote}

Comfort was therefore shaped by bodies, family routines, climates, and past living environments, which influenced what forms of energy use felt reasonable.

\subsubsection{Care and Safety}

Care and safety provided another basis for treating high consumption as legitimate in certain situations. Taiwan's particularly humid climate made energy-intensive dehumidification systems an unavoidable choice for many families. Twelve parents saw dehumidification as a necessary household need, and families used combinations of dehumidifiers, drying systems, and air purifiers to keep their homes dry. Melissa described her ongoing struggle with humidity:

\begin{quote}
\textit{``The dehumidifier uses the most electricity. Air conditioners, heaters, and the bathroom heater all use a lot too... But in Taiwan's humid climate, you cannot really avoid using them. If you do not use a dehumidifier, your house stays damp. If you do not use the bathroom heater, mold starts growing, and everything turns yellow. That is just not the life we want.''}
\end{quote}

Melissa's care for the family's living environment made the high energy use of dehumidification legitimate. It protected the home from mold and discoloration and helped maintain an acceptable living environment for the family. Other families also accepted additional consumption for specific care needs, such as safety. In the Wilson family (F14), Jessica kept the staircase lights on when her mother-in-law stayed with the family so she could move safely through the home at night.

Sometimes care needs were temporary, yet families still treated the resulting consumption as necessary. During COVID-19, Jennifer in the Johnson family (F2) replaced a smaller refrigerator with a larger double-door model because the household needed to store more food at once. She acknowledged its higher energy use: \textit{``The government labels appliances by energy-consumption level. Our refrigerator is in the highest tier.''} Even so, she did not regret the purchase because the family genuinely needed greater food storage at the time. 

In these cases, appropriate consumption depended on the relationship or need that energy sustained: a child who could sleep without fear, an older relative who could move safely through the home, or a household that could store enough food. Care shaped what families considered necessary and acceptable energy use in everyday life.

\subsubsection{Fun and Pleasure}

Fun raised another conflict of energy use in family life. Fourteen parents regarded children's long showers as \textit{``a waste of water''.} They played with water, sang in the bathroom, and lingered after washing was complete. The Hall family (F11) shows parents' and children's diverse understanding of shower time. Sometimes Amanda thought Ethan had already finished showering and went to check on him, only to find him still singing and playing with water, having completely lost track of time.

\begin{quote}
\textit{``He just really enjoys having nothing on and having the water rush over his body. He feels very comfortable. He even uses `cleaning the bathroom' as an excuse to stay inside a little longer [...] This is his happy time, but I am standing outside getting angrier and angrier.''}
\end{quote}

Ethan described the same period as absorbing enough that he stopped noticing time: \textit{``I get so absorbed in singing... I do not think about the time.''} For Ethan, the bathroom offered a private space for relaxation and play. Other children similarly described hot water as bodily comfort or play. Some kept water running while soaping because turning it off made them cold. Others only realized how long they had been showering after a parent called through the door. Angela said that her child \textit{``does not know how long ten minutes is.'}

Fun was not unique to children as an energy-related need. Adults also found pleasure from energy use. Brandon in the Smith family had filled the home with sensors, linked appliances, one-touch scenes, automated balcony watering, and connected AC and fan controls. Ashley saw this system largely as her husband's hobby and did not share his enthusiasm:

\begin{quote}
\textit{``He really enjoys being able to turn everything on or off together with one button. It's quite fun for him. But I think all of those devices also consume electricity... Our home is quite small. It is not Bill Gates's house. We do not need so many one-touch functions.''}
\end{quote}

Brandon acknowledged that tinkering with smart-home devices was a pleasure for him, but still emphasized that it could bring energy savings. Broadly speaking, participants treated energy use for fun as more personal and less legitimate than shared goals such as thrift. Thus they often framed it as a ``bad habit'' to be corrected, such as children taking overly long showers.

\subsection{Adaptation: Continually Making the \textit{Good} Life Work}

As noted in Section 5.1, participants shared thrift as a common value in energy use, while often differing in other values that shaped a \textit{good} family life, related to their upbringing, embodied experiences, and local environment. In this section, we examine how families made a \textit{good} family life work through continuous  \textit{adaptations} to everyday energy practices and smart-home arrangements as circumstances changed and tensions emerged. These practices were embedded in existing divisions of household labor, family relations, and wider sociopolitical contexts.

\subsubsection{Adapting to Changing People and Routines}

Household energy rules were usually maintained through repeated reminders rather than formal agreement. Nicole laughed when asked how the Clark family discussed energy use: \textit{``We would not hold a family meeting specifically to discuss how to save energy.''} Expectations instead appeared in everyday reminders to turn off lights and AC, shorten showers, or switch off unused appliances. This reminder work was frequent and often tiring. Melissa described it as continuous maternal labor: \textit{``Reminding children happens all day, every day... Isn't that just what mothers do?''} Ashley in the Smith family similarly explained, 

\begin{quote}
\textit{``Getting children to remember is quite difficult... They keep forgetting to turn off the lights or the AC, so I still have to keep nagging them. Even I find it annoying, but that is just what being a mother is like.''}
\end{quote}

Across our participants, mothers more often took on repeated reminder work, while fathers more often configured household energy technologies and introduced automation intended to ``resolve'' this repetition. Ten fathers used automated AC schedules in children's bedrooms. Brandon adopted smart AC remotes connected to a mobile app on his phone. He configured an AC schedule for Emma's room, but Emma did not fall asleep at the same time every night. Ashley therefore waited until Emma actually fell asleep before manually activating the auto-off function:

\begin{quote}
\textit{``Because my child does not go to bed at the same time every day, I have to manually activate the AC's auto-off schedule after she falls asleep. But I often forget, and then the next morning my husband finds that the air conditioner is still running and reminds me on LINE.''}
\end{quote}

A similar mismatch appeared in the Wilson family. Jessica found that a fixed shutoff time did not fit Charlotte's bodily response to the room temperature: \textit{``As soon as the AC turns off, she notices. It does not even take five minutes. She notices, and then she starts crying.''} Jessica therefore adjusted the AC around when Charlotte actually woke. In both families, automation depended on someone noticing when the child's routine or comfort no longer matched the anticipated schedule and manually adapting the arrangement.

Changes in who was present in the home could also disrupt established routines. In the Martin family, Isabella and her parents used schedules to control the AC, while her grandmother did not speak English and preferred physical switches. Isabella explained: \textit{``When Grandma comes over, she does not speak English, and she is not used to these things. She has to turn things on or off by hand. Sometimes we only realize the AC did not turn on automatically after the room gets hot, because Grandma switched it off by hand last time.''} Reminders, schedules, and smart-home configurations could stabilize routines, but keeping them workable still required family members to notice changes and adapt the arrangement when those routines no longer fit.

\subsubsection{Adapting to Nearby Care Needs}
To pursue a \textit{good} life, families made adjustments and exceptions in energy use to adapt to care needs. These adaptations could require fine-grained and ongoing adjustment. Melissa managed this between two sons who experienced cooling differently. One was especially sensitive to heat, while the other became cold easily. She combined AC settings, blankets, and a fan instead of treating one temperature as appropriate for both children:

\begin{quote}
\textit{``The older one gets hot very easily, and the younger one gets cold very easily. The older one still uses a thin blanket with the AC, but the younger one needs a thick blanket even in summer. I want both of them to sleep well, so I cannot set the temperature too low... Sometimes the older one says he is hot and points the fan at himself. After he falls asleep, I turn the fan down and point it toward the floor.''}
\end{quote}

These fine-grained adaptations often depended on family relationships and intimate knowledge of one another. Melissa understood why one son turned on multiple lights when entering a large bedroom at night because he was afraid of the dark. In the Anderson family (F15), Stephanie kept the living-room television on so her son Elijah, who was afraid of the dark, could sleep with his bedroom door slightly open and sense that family members were nearby.

\subsubsection{Adapting to Material and environmental conditions}

Families also adapted energy practices to Taiwan's climate and the material conditions of their homes. Avoiding heating was described by eight participants as a distinctly Taiwanese way of saving energy. Some chose not to use or even install heating because they saw it as energy-intensive and found heated air uncomfortable. Amanda explained: \textit{``In winter, I actually don't think we need the heater. The coldest weather only lasts about a week. The heated air also makes my throat uncomfortable.''} Instead, her family got dressed in the bathroom after showering to make use of the remaining warmth before entering the bedroom, avoiding the need to turn on the heater. Although she chose not to install heating, she felt that dehumidification was unavoidable, since her family lived in one of the most humid places in Taiwan:

\begin{quote}
\textit{``The renovation company initially refused to install a dehumidification system because our walls are not concrete, so they were worried the system would be too heavy [...] But I did need a dehumidifying and clothes-drying system. It has a drying rack with a heating function that blows warm air from above. I have to use the warm air together with a dehumidifier to get the clothes dry. There are five people in our family. In winter, each person may wear four or five pieces of clothing, so that is around twenty pieces in total. If today's clothes do not dry, there will not be enough space for the clothes we change out of the next day.''}
\end{quote}

Rachel's bathroom routine in the Taylor family (F17) shows how families worked within the humid environmental and material constraints. Her bathroom had no exterior window and poor ventilation, so moisture accumulated after family members showered. Rachel ran the dehumidifier after everyone had finished and periodically used a UV light for sterilization. Because UV could harm skin, she timed its operation around when nobody would enter the room:

\begin{quote}
\textit{``The bathroom ventilation is not very good, so after everyone finishes showering, we run the dehumidifier for about half an hour to an hour... Dehumidifying is my task because it is usually late, after everyone has showered. The UV light is also my responsibility because it is more complicated... UV can hurt your skin, so I turn it on when nobody needs to use the bathroom, sometimes before we go out, and then I turn it off when we come back.''}
\end{quote}

In those cases, adapting energy use meant continually working within the material limits of the home and local climate, while coordinating technologies around family routines, safety, and care needs to keep the family infrastructure functioning well.

\subsubsection{When Machines Breakdown}
Families also adapted when smart-home infrastructures stopped working as expected. Tyler had connected many devices in the Thomas family (F16) through Xiaomi's ecosystem. When the network failed, the convenience of this arrangement disappeared abruptly:

\begin{quote}
\textit{``Once the network went down, these things instantly became stupid. Mom would complain that they did not work... In that situation, manual control is still better. You can still use things manually without the network, but once people get used to automation, suddenly switching back to manual feels unfamiliar. After a long time, you can even forget that you need to do it manually.''}
\end{quote}

The breakdown exposed how household routines had been reorganized around connectivity. Family members had to remember older forms of control and determine which functions still worked. Other failures required family members to notice abnormal behavior during everyday use. Angela in the Jackson family (F19) discovered a malfunctioning smart toilet only after getting up during the night. The water kept running because the tank could no longer fill properly, and the problem later appeared as an unusual increase in the household's water bill. She recalled:

\begin{quote}
\textit{``One time it was the smart toilet. I don't know why, but the water kept running. It seemed to have malfunctioned, and the tank could not fill up. At first, we could not figure out where the leak was coming from or why the water bill was so high. Then one night, I got up and heard the sound of running water.''}
\end{quote}

Heather in the Moore family (F18) encountered a similar problem with the family's energy-efficient heat-pump water heater. The system normally ran on a preset schedule and required little attention. When Heather needed hot water outside the usual schedule, she noticed that the control panel was no longer responding properly. She turned to her husband for help, but he couldn't fix it. They finally contacted a service technician. The adaptation work, including noticing and seeking repair, became visible when the system stopped functioning as expected.

\subsubsection{Parent--Child and Intergenerational Adaptation}

Children were also active participants in adaptation. Parents often described them as forgetful and in need of reminders, yet children also corrected adults, formed alliances, and became technical intermediaries. In the Hall family (F11), Amanda noticed that Ethan resisted reminders from his older sister and told him: \textit{``In this family, if you do something wrong, someone will call you out, no matter how old or young you are.''} Ethan later began reminding his mother to turn off the lights. Mia in the Clark family (F13) similarly became known as the household's \textit{AC controller}. After asking Nicole to turn off the AC before leaving home, Mia returned to find it still running and told her: \textit{``Mom, you did not turn off the AC this morning. You are wasteful too!''}

Nicole initially treated these reminders as part of ordinary family interaction, but one incident showed how consequential a child's attention could become. Her husband had left a cast-iron pan drying over a low flame on the stove and forgotten about it. The stove remained on overnight until Mia noticed it the next morning:

\begin{quote}
\textit{``There was one really scary incident. We have a cast-iron pan that needs to be dried after washing. Dad put it over the lowest flame on the stove to dry and then walked away, as he usually does. None of us noticed, and it kept burning until the next morning. When we were about to leave, luckily my daughter looked over and said, `Why is there still a flame on the stove?' I went over and turned it off. That was when I realized, my goodness, it had never been turned off.''}
\end{quote}

Afterward, Nicole and Mia worked together to supervise Mia's father not to leave the stove on while drying the pan. Children also formed alliances with one parent to challenge another family member's energy use. In the Jones family (F5), Samantha and Noah worked together to convince his father to stop setting the AC unnecessarily low. Samantha recalled: \textit{``We even did an experiment together for his dad, to show him that setting the AC too low only makes the machine keep running. The room cannot actually get that cold.''}

Children could also become technical intermediaries. In the Lee family (F21), Kevin introduced many of the family's smart-home devices, while his eight-year-old son Henry became interested in exploring their functions through Kevin's phone. Over time, Henry became the person family members asked when a device did not respond. Google Home made this role especially visible. Henry explained: \textit{``I only have to say it once and it turns on... I am the first person Google understands, my sister is second, Dad is third, and Mom is fourth.''} Megan confirmed what happened when Google ignored her: \textit{``I just ask him to say it, and he takes care of it.''} Abigail also noted that speaking too quickly or with a hoarse voice could cause the assistant to ignore her. Henry's role therefore emerged from both his interest in the system and its unequal responsiveness to different family members.

Intergenerational relationships further complicated adaptation. Four families described grandparents intervening in household energy practices, often bringing different understandings of thrift and different habits into the home. Rachel, for example, found it difficult when her parents visited and resisted using enough air-conditioning because they wanted to save energy. Melissa described how a grandmother could unintentionally interrupt her efforts to teach her children energy-saving habits:

\begin{quote}
\textit{``Sometimes my younger son's desk lamp or night-light is left on, so I remind him to turn it off. If they still do not turn the lights off, I will do it, but then I punish them by not letting them turn on the lights at night. They are very scared of that. When Grandma is here, she will turn the lights off for them. I do not think she knows about the punishment. She is simply more frugal and habitually switches them off as she passes by. But I think teaching the children to develop good habits is more important.''}
\end{quote}

Across these families, energy rules did not move only from parents to children. Children reminded, corrected, and sometimes worked with adults, while grandparents brought another generation's habits into the household. Adaptation therefore involved ongoing negotiation across generations over who should notice, remind, intervene, and take responsibility for household energy practices.

\subsection{Co-designing Future \textit{Good Lives}}

The co-design sessions made these lived positions visible in participants' future smart-home imaginaries. Participants did not approach design as abstract users with the same goal. Those who spent more time configuring systems often imagined stronger automation. Those who performed repeated reminders or adjustments imagined technologies that could share that work. Children often imagined reducing parental intervention or gaining more direct access to household controls.

\subsubsection{Automating Away Negotiation}

Several fathers imagined automation as a way to reduce repeated communication. Tyler stated this aspiration most directly: \textit{``The best way is no communication. I just use smart devices.''}

His proposal grew from an existing practice. Tyler already linked window sensors, fans, lighting, and AC controls so that one action could trigger another. He wanted these relationships to operate without requiring someone to remember a command or negotiate each use. Other fathers proposed motion-triggered lighting, automated AC schedules, remote shutoff, and linked whole-home scenes. The attraction was clear: if a device could detect an empty room and switch itself off, nobody needed to notice the waste or confront another family member about it. Yet even P8, an avid smart-home enthusiast, acknowledged that highly context-dependent systems such as lighting were difficult to automate reliably.

\subsubsection{Delegating Reminder Work}

Parents who performed repeated reminder work often imagined technology taking over the act of reminding. The Jones family's co-design session made this desire especially clear. Samantha had already spent years negotiating her husband's low AC settings and reminding children about household energy use. She imagined centralized controls that could enforce an acceptable temperature range, while Noah proposed a robot that could intervene when the AC needed to be adjusted:

\begin{quote}
\textit{Samantha: ``Wouldn't the robot be more effective than Mom reminding you?''}

\textit{Noah: ``Yes... because it is not a person.''}
\end{quote}

Samantha interpreted his answer as a way to reduce the interpersonal tension of repeated reminders: \textit{``Because it doesn't nag. It is already set, so there isn't that nagging problem.''} Children also imagined technologies that could remind them directly. In the Lee family, Abigail designed a wearable for keeping track of time in the shower:

\begin{quote}
\textit{``You could wear a watch while you shower. It could remind you every twenty minutes. If you still don't come out, it would start screaming.''}
\end{quote}

Across these designs, parents and children imagined technical reminders as a way to reduce repeated calling, nagging, and face-to-face enforcement. The family rule remained, but the act of reminding could be shifted from one family member to a device.

\subsubsection{Centralizing Control and Making Consumption Visible}

Some mothers imagined future systems that would make the dispersed work of household coordination easier to see and act on. Seven proposed centralized interfaces for viewing or managing appliances across rooms. Heather, a stay-at-home mother, noted that her husband had introduced most of the family's smart-home devices: \textit{``He's the one who introduced it, so he is in control, and I just use them.''} This arrangement became inconvenient when she needed to change settings herself. Adjustments such as modifying the AC schedule had to be made through her husband's phone, leaving her dependent on him for even routine changes.

Others wanted device-level energy data that could identify which appliances were responsible for unusually high bills. Ashley's design in the Smith family grew directly from recurring disagreement with Brandon about the source of their high electricity use. Brandon compared current bills with previous periods and emphasized reducing consumption. Ashley suspected that public-area electricity charges or the many Xiaomi devices Brandon had installed could also contribute, but neither had enough device-level evidence to resolve the disagreement. She therefore sketched an app that would make consumption visible by appliance and location:

\begin{quote}
\textit{``The best would be an app that could detect the electricity use of every appliance. It could separate them by location, or I could switch the view and group them by appliance... Then I could compare how much that appliance used in the previous period. You could easily see which room is using more electricity, or which appliance is actually using much more than you expected.''}
\end{quote}

Ashley also imagined using this evidence in family discussions. If a child's room consumed more than comparable rooms because a light was repeatedly left on, she felt the data would give her a concrete reason for explaining the problem. The same logic applied to disagreements with Brandon. Instead of arguing from intuition about which devices were responsible for the bill, family members could point to a shared record. The design therefore connected visibility with negotiation. Ashley did not imagine data only as feedback for an individual optimizer. She wanted evidence that could enter a family conversation about whose devices, routines, and expectations were producing consumption.

The Lee family's (F21) co-design extended this concern from centralized control to who could actually access it. In everyday use, Google Home recognized Henry's voice most reliably, while Megan sometimes relied on him to issue commands for her. Henry responded to this uneven access by imagining support for more family members: \textit{``I want Google Home to recognize ten voices. Right now, we only have two set up, so everyone has to ask me to turn things off.''}

For Henry, improving the smart home meant expanding who could reliably act through it, rather than simply centralizing more functions. These proposals show that making household energy use easier to manage involved both making consumption visible and making control accessible across family members.

\subsubsection{Questioning the Limits of Automation}

Participants' designs also exposed doubts about whether smart-home systems could fully anticipate family life. A child's bedroom needed different lighting while studying, sleeping, playing, or waking at night. A bathroom needed ventilation and dehumidification after showering but could not safely run a UV light while someone was inside. An AC schedule could fit a usual bedtime and fail on the night a child stayed awake later.

Nicole from the Clark family described the deeper problem as having too many people trying to direct the same household system:

\begin{quote}
\textit{``When everyone wants to control the same device, who should it listen to? Technology can only help. First, we have to agree on things like when it's appropriate to use the air conditioner, what temperature feels comfortable, or whether children should decide the temperature in their own rooms. We live together, so someone always has to compromise.''}
\end{quote}

Her account places a limit on the idea that better sensing alone can resolve household energy tensions. A motion sensor can detect movement without knowing why someone entered a room. A schedule can reproduce a typical bedtime without knowing that a child is still awake. A voice assistant can execute a command while responding more reliably to Henry than to Megan. These technologies can act on household conditions, but families still interpret whose comfort and needs matter in that moment. 
\section{Discussion}

Smart-home sustainability often begins from an imagined household good: a family identifies desirable energy practices, configures technologies around them, and expects automation to help sustain those practices. Our findings show that this ``good'' took two different forms in everyday family life. Drawing on Rousseau, we use \textit{collective good} to describe values around which family members could broadly orient together. In our households, thrift and the expectation of \textit{not wasting} came closest to this form of agreement. Drawing on Mouffe and DiSalvo, we use \textit{plural and contestable good} to describe legitimate positions that remained unsettled within the family \cite{rousseau1997social,mouffe2005political,adversarialdesign}. Comfort, safety, and enjoyment often took this form because family members experienced the same temperature, schedule, shower, or lighting arrangement differently.

This distinction helps us ask different questions of family IoT. For plural and contestable good, the design problem concerns whose position becomes encoded as the household default and whether others can continue to contest it. For collective good, the design problem concerns how values such as thrift are learned, practiced, and reinterpreted as family life changes. Both depend on continuing adaptive work. We first discuss how family IoT can keep plural good open to contestation. We then consider how technology might support the intergenerational articulation of collective good such as thrift. Finally, we examine the adaptive care work through which families sustain both forms of good as bodies, routines, technologies, and relationships change.

\subsection{Keeping \textit{`Plural Good'} Open to Contestation}

\subsubsection{When One Position Becomes the Household Default}
Our findings show that comfort, safety, and enjoyment produced differences that families could not settle once and for all. Nicole (F13) described the basic problem as ``one person feels hot and another feels cold,'' while Amanda (F11) saw Ethan's long showers as excessive consumption and Ethan experienced them as his ``happy time.'' These positions could coexist without converging on a single definition of appropriate energy use. Rousseau's discussion of the family sharpens the issue of authority and consent: parental authority is tied to children's dependence, while continued association beyond dependence rests on convention \cite{rousseau1997social}. Yet once temperatures, schedules, or routines are encoded into smart-home systems, a personal preference can acquire the appearance and durability of a household decision. In the Smith family (F1), Brandon built one-button scenes that Ashley questioned; in the Thomas family (F16), Tyler imagined that ``the best way is no communication'' because smart devices could carry out decisions automatically.

The central design problem is therefore who can translate a personal preference into an automated household rule, and whether others can still question it. This authority was also gendered. Bell and Dourish's shed lens shows how domestic technologies enter spaces already organized through gendered differences in programming and control \cite{bell2007shed}. Tyler sometimes disabled the wall control connected to the family's Xiaomi lighting so his wife could not use XiaoAi to turn on the lights in the morning; she interpreted the assistant as broken and had to operate the light herself. Access could also be uneven at the interface level. In the Lee family (F21), Google Home responded most reliably to eight-year-old Henry and least reliably to his mother Megan, who often asked Henry to repeat ignored commands. Smart-home systems can therefore make one member's position durable while making others dependent, turning technical access into household authority. Design should preserve opportunities to contest and revise such defaults.

\subsubsection{Designing to Keep Household Differences Contestable}
Family technology research has long documented differing expectations, conflicts, and negotiations among household members \cite{blackwell2016managing,derix2022family}. Our findings extend this work by showing that disagreement can itself be productive. Mouffe's agonistic pluralism treats legitimate disagreement as an enduring feature of collective life \cite{mouffe2005political,mouffe1999deliberative}, while Sprey similarly understands family life as ongoing conflict over shared resources and arrangements \cite{sprey1979conflict}. Children actively participated in these negotiations: Ava (F4) directly challenged her mother by saying, ``Mom, you are wasteful too,'' and Mia (F13), the family's ``AC controller,'' called out Nicole for forgetting to turn off the AC. Following Coser, such arguing, correcting, and overriding can help members reaffirm and revise shared values rather than simply disrupt family life \cite{coser1956functions}. Family IoT should therefore not treat disagreement as noise to eliminate, but preserve meaningful opportunities for family members, including children, to challenge how household arrangements are defined and enforced.

Adversarial design provides one way to support this contestability. Following DiSalvo, household settings can be treated as positions that remain visible and open to challenge \cite{adversarialdesign}. Usman Haque's \textit{Natural Fuse}, for example, links plant units, appliances, and users so that drawing beyond a local energy allocation in ``selfish'' mode reduces resources available to others, materializing relations among personal desire, shared resources, and consequences \cite{adversarialdesign}. Recent work on smart-home control and observability offers technical starting points for similar designs \cite{xue2024control,karanika2026rasc}. Observability could extend from device execution to the social provenance of a setting: a cooling rule might show whose comfort it serves, a reminder whose concern it carries, and an override who changed a setting and why. Defaults could require periodic re-affirmation rather than persist silently. The aim is not to eliminate Nicole's ``too many horses steering the same cart,'' but to make visible which horse is currently pulling and allow others to redirect it.

\subsection{Supporting Collective Good Across Generations}

\subsubsection{Re-articulating Thrift Across Generations}
Thrift presented a different design problem because \textit{not wasting} was widely shared across generations. Ten parents traced practices such as switching off unused lights and saving water to lessons learned from their own parents. Nicole (F13) similarly described herself and her husband as belonging to a generation taught to avoid waste. Yet the meaning of thrift had shifted. Earlier generations often tied conservation closely to household expenses, whereas many parents in our study described electricity and water as relatively affordable. Michael (F3) captured what remained: ``Just use energy normally and don't waste it.'' Conservation was therefore less about minimizing every unit of consumption than distinguishing reasonable use from waste. Repeating practices such as switching off a light allowed families to carry an inherited value into new material conditions while teaching what responsible use of shared resources meant. A collective good such as thrift was sustained not only through agreement, but through repeated domestic practice and mutual accountability.

This intergenerational role can disappear when automation is judged only by energy outcomes. In the Smith family (F1), Ashley described repeatedly reminding Emma to switch off forgotten lights as exhausting, while Brandon said, ``If they learn to turn off the lights when they leave, that is good enough.'' An automated light could achieve the same immediate reduction without requiring Emma to remember or respond, whereas the reminder made conservation something practiced within a family relationship. Value Sensitive Design highlights values embedded in technologies, and values-as-lived-experience work emphasizes how values emerge through situated practice \cite{friedman2006value,ledantec2009values}. Our findings extend this concern to the intergenerational life of values. Slow Technology \cite{hallnas2001slow} and later work on reflection, participation, and evolution \cite{asadi2023calming} suggest preserving some moments of attention rather than automating them away. Reminders, visualizations, or delayed automation could create opportunities to reflect on why a resource matters, how an expectation developed, or when an exception is justified.

\subsubsection{Supporting Children as Participants in Family Values}
The intergenerational transmission of thrift also raises what role children should have in interpreting shared family values. Parents often described successful energy education through children's compliance with reminders and household rules, echoing Kohlberg's account of moral reasoning organized around interpersonal approval and established social rules \cite{kohlberg1969stage}. Gilligan's critique instead foregrounds moral understanding that develops through concrete relationships, communication, and care \cite{gilligan1977different,gilligan1982different}. Children in our study already moved beyond simple compliance. Ava (F4) challenged her mother for wasting energy, Mia (F13) monitored the AC and held Nicole accountable, and Amanda (F11) told her children that ``if you do something wrong, someone will call you out, no matter how old or young you are.'' These cases show children not merely receiving family rules, but participating in interpreting, applying, and enforcing them in everyday family life. Collective good is therefore sustained through intergenerational participation rather than one-directional transmission alone.

Family IoT could support children as active participants in this process. Abigail's co-design idea in the Lee family (F21) illustrates one possibility: she imagined a wearable that would remind her every twenty minutes while showering, shifting the reminder from a parent outside the bathroom to something she could notice and respond to herself. Reminder robots, timers, or automatic shut-offs could similarly communicate household expectations while still making visible why a rule exists, whose concern it represents, and when comfort or safety warrants an exception. Children could respond, propose exceptions, or question adults when expectations are applied unevenly. This matters because automation can change the cultural work carried by an everyday action even when the measurable energy outcome remains identical. Designing around collective good should therefore preserve selected opportunities for teaching, reflection, and participation, allowing children to help reinterpret inherited values as family circumstances change rather than treating conservation as either parental instruction or invisible automated behavior.

\subsection{Sustaining Family Goods Through Adaptation and Care}

\subsubsection{Adaptation as Situated Care Work}
Shared and contestable family goods depended on continuing adaptive care work through which families noticed changing conditions and reworked energy arrangements. Tronto's care ethics helps specify this work as attentiveness, responsibility, competence, and responsiveness \cite{tronto1993moral}. Ashley in the Smith family waited until Emma actually fell asleep before activating the AC's auto-off function, while Samantha in the Jones family raised the temperature after noticing her children tightly wrapped in blankets. Xiang's notion of the \textit{Nearby} further directs attention to the concrete relations among bodies, routines, rooms, appliances, weather, and material conditions that made such adjustments necessary \cite{xiang2021nearby}. Haraway's situated knowledges and Kempton's account of household energy reasoning likewise show that people know what needs changing from particular embodied and practical positions \cite{haraway1988situated,kempton1986two}. Adaptation is therefore not generic flexibility, but situated care: noticing when an arrangement no longer fits, interpreting why that matters for particular people and conditions, and revising it accordingly.

\subsubsection{The Uneven Distribution of Adaptive Care}
This adaptive care was also unevenly distributed. Tronto's concept of privileged irresponsibility helps explain how some family members can benefit from care while remaining less aware of the work that sustains it \cite{tronto1993moral}, giving empirical specificity to the imaginaries of \textit{Resource Man} and the \textit{Smart Wife} \cite{resourceman,smartwife}. Fathers more often configured schedules, integrated devices, and pursued technical optimization, while mothers more often noticed when those arrangements no longer fit routines, bodies, or care needs and made the necessary adjustments. Material care formed part of this work as well: families operated dehumidifiers, detected appliance faults, worked around infrastructural limitations, and maintained homes against humidity and mold, consistent with accounts of care as maintaining relations with objects and ongoing attunement to changing situations \cite{puig2017matters,Attuning}. These different positions also shaped future imaginaries: fathers more often proposed automation that settled decisions in advance, mothers emphasized coordination and reminders, and children imagined greater access and independence.

\subsubsection{Designing for Adaptation, Not Seamlessness}
For design, adaptability therefore means more than adding flexible settings to automation. Dourish describes context as continually produced and negotiated through practice \cite{dourish2004context}, while Davidoff et al. argue that smart homes must accommodate multiple, overlapping, and sometimes conflicting domestic goals \cite{davidoff2006principles}. Our findings add attention to the relational labor required to make such flexibility possible. Family IoT should help members notice when bodies, routines, care needs, technologies, or interpretations of waste change; understand whose needs are implicated; make exceptions; and revise existing arrangements. This also changes how smart-home success should be evaluated. A system that appears seamless may simply shift adaptive work onto the person who continues to notice what automation misses. Designs that settle plural goods can privilege one member's position, while designs that fully automate collective goods can reduce opportunities for values to be practiced and reinterpreted. Sustainable family IoT should therefore support adaptation while making visible what is settled, what remains open, and who sustains the arrangement.
\section{Limitations and Future Work}


Our study focuses on heterosexual parent-child families, mostly from northern Taiwan and with relatively high household incomes. This sample was partly shaped by the time and trust required for in-home research. Future work could include more diverse family forms, such as single-parent, same-sex, and blended families. These settings may reveal different patterns of household authority, care work, and gendered divisions of energy-related labor. We also examined energy practices at one point in time and could not trace how family negotiations change across seasons, children's development, or longer-term use of smart-home technologies. Longitudinal work, such as diary studies or repeated interviews, could examine how these arrangements evolve over time. Our analysis centered on everyday negotiation, care, and adaptation, leaving domestic governance as an important direction for studying authority, rule-making, and participation within smart homes. Future work could also extend this household-level perspective to smart-city infrastructures and examine how family energy negotiations interact with wider systems such as smart meters, utility platforms, and municipal sustainability programs.

\section{Conclusion}

In this paper, we offer sustainable HCI a family-centered understanding of how parent-child families negotiate and sustain a \textit{good} family life through everyday energy practices. Drawing on in-home interviews and co-design sessions with 21 Taiwanese families (46 parents and children), we show how judgments about thrift, comfort, care, safety, and enjoyment shaped what families considered appropriate energy use, and how these arrangements required continual adaptation as people, routines, technologies, and material conditions changed. We distinguish between ``collective good'' and ``plural and contestable good'' which family members experienced differently. Drawing on the Nearby, care ethics, and adversarial design, we conceptualize adaptation as situated care work that was unevenly distributed across family members. Our work contributes design directions for family IoT that support the continued articulation of collective values, keep plural claims visible and open to contestation, and recognize the adaptive work through which families keep everyday arrangements workable.




\bibliographystyle{ACM-Reference-Format}
\bibliography{citation}

\appendix

\clearpage

\begin{table*}[t]
\section{Codebook}
\centering
\caption{Codebook developed from the reflective thematic analysis.}
\label{tab:codebook}

\scriptsize
\setlength{\tabcolsep}{3.5pt}
\renewcommand{\arraystretch}{1.1}

\begin{tabularx}{\textwidth}{
@{}
>{\raggedright\arraybackslash}p{0.045\textwidth}
>{\raggedright\arraybackslash}p{0.39\textwidth}
>{\raggedright\arraybackslash}X
@{}
}
\toprule
\textbf{Code} & \textbf{Code and Sub-codes} & \textbf{Description} \\
\midrule

\multicolumn{3}{
@{}>{\raggedright\arraybackslash}p{\textwidth}@{}
}{
\textbf{RQ1: How do parent-child families make sense of ``good'' family life through everyday home energy practices in Taiwan?}
} \\
\midrule

1 &
\textbf{Making Sense of Appropriate Energy Use}

\textit{Sub-codes:}
Avoiding waste and practicing thrift;
Balancing efficiency and necessary use;
Legitimizing comfort, care, safety, and enjoyment
&
How family members judge whether energy use is appropriate, excessive, or necessary.
These judgments draw on thrift, cost, efficiency, bodily comfort, care needs, safety, and enjoyment.
The same use can be understood differently by different family members.
\\[0.5em]

2 &
\textbf{Learning and Contesting Family Energy Values}

\textit{Sub-codes:}
Passing down family values;
Learning through rules and reminders;
Holding others accountable;
Challenging or reinterpreting rules
&
How energy values become part of family life across generations.
Parents teach practices such as turning off lights or limiting water use, while children can follow, question, or apply those expectations back to parents and siblings.
\\

\midrule

\multicolumn{3}{
@{}>{\raggedright\arraybackslash}p{\textwidth}@{}
}{
\textbf{RQ2: How do families adapt energy practices and smart-home arrangements as household needs and circumstances change?}
} \\
\midrule

3 &
\textbf{Recognizing When Existing Arrangements No Longer Fit}

\textit{Sub-codes:}
Changing routines and schedules;
Changing bodily and care needs;
Environmental and material conditions;
Technology mismatch or breakdown
&
Moments when an existing energy rule, setting, or automated arrangement stops fitting everyday life.
Change may come from irregular routines, different comfort needs, humidity and housing conditions, or devices that fail technically or socially.
\\[0.5em]

4 &
\textbf{Adapting Energy Arrangements in Practice}

\textit{Sub-codes:}
Reminding and monitoring;
Adjusting and overriding settings;
Making exceptions and compromises;
Working around or repairing technology
&
Concrete actions family members take to keep household arrangements workable.
Adaptation includes noticing problems, reminding others, changing settings, allowing exceptions, negotiating compromises, and falling back to manual practices when needed.
\\[0.5em]

5 &
\textbf{Distributing Control and Adaptive Work}

\textit{Sub-codes:}
Introducing and configuring technology;
Managing everyday use and maintenance;
Children as monitors or technical intermediaries;
Uneven control and responsibility
&
Who has authority over devices and who performs the ongoing work required to sustain them.
Initial setup and technical control can differ from daily monitoring, adjustment, reminder, and repair work, creating gendered and generational differences in responsibility.
\\

\midrule

\multicolumn{3}{
@{}>{\raggedright\arraybackslash}p{\textwidth}@{}
}{
\textbf{RQ3: How do family members' experiences of this adaptive work shape their imaginaries of future smart-home technologies?}
} \\
\midrule

6 &
\textbf{Imagining Technology to Reduce Repetitive Work}

\textit{Sub-codes:}
Automating routine actions;
Delegating reminders to devices;
Centralizing control and monitoring
&
Future designs imagined in response to repeated reminders, forgetting, fragmented controls, and everyday coordination work.
Participants often proposed automation, timers, reminder robots, or centralized interfaces to reduce repeated household labor.
\\[0.5em]

7 &
\textbf{Imagining More Flexible and Participatory Smart Homes}

\textit{Sub-codes:}
Expanding access and control;
Supporting child self-regulation;
Keeping manual override and exceptions;
Making energy use and decisions visible
&
Imaginaries shaped by the limits of current smart-home arrangements.
Participants envisioned technologies that give more family members direct access, support children in managing their own practices, preserve human judgment, and make settings or energy use visible enough to support discussion and revision.
\\

\bottomrule
\end{tabularx}
\end{table*}

\end{document}